%% file: 0.main.tex
\documentclass[lettersize,journal]{IEEEtran}
\usepackage{amsmath,amsfonts}
\usepackage{algorithmic}
\usepackage{algorithm}
\usepackage{array}
\usepackage{textcomp}
\usepackage{stfloats}
\usepackage{url}
\usepackage{verbatim}
\usepackage{graphicx}
\usepackage{cite}
\usepackage{subimages}
\usepackage[usenames]{color}
\usepackage{ulem}
\usepackage[font=small,labelfont=bf]{caption} 
\newcommand{\ordena}{ORDENA }

\newcommand{\secref}[1]{Section~\ref{sec:#1}}

\newcommand{\myparagraph}[1]{\noindent\textbf{#1}}
\usepackage{xcolor}

\let\originaltextcolor\textcolor

\renewcommand{\textcolor}[2]{%
  \originaltextcolor{black}{#2}%
}

\begin{document}

\title{\ordena\!\!: ORigin-DEstiNAtion data exploration}
\author{Karelia Salinas, Victor Barella, André Luiz Cunha, Gabriel Martins de Oliveira, \\ Thales Viera, and Luis Gustavo Nonato
}

\markboth{Journal of \LaTeX\ Class Files,~Vol.~14, No.~8, August~2021}%
{Shell \MakeLowercase{\textit{et al.}}: A Sample Article Using IEEEtran.cls for IEEE Journals}


\maketitle

\begin{abstract}
Analyzing origin-destination flows is an important problem that has been extensively investigated in several scientific fields, particularly by the visualization community. The problem becomes especially challenging when involving massive data, demanding mechanisms such as data aggregation and interactive filtering to make the exploratory process doable. However, data aggregation tends to smooth out certain patterns, and deciding which data should be filtered is not straightforward. In this work, we propose \ordena\!\!, a visual analytic tool to explore origin and destination data. \ordena is built upon a simple and intuitive scatter plot where the horizontal and vertical axes correspond to origins and destinations. Therefore, each origin-destination flow is represented as a point in the scatter plot. How the points are organized in the plot layout reveals important spatial phenomena present in the data. Moreover, \ordena provides explainability resources that allow users to better understand the relation between origin-destination flows and associated attributes. 
We illustrate \ordena\!\!'s effectiveness in a set of case studies, which have also been elaborated in collaboration with domain experts. The proposed tool has also been evaluated by domain experts not involved in its development, which provided quite positive feedback about ORDENA.

\end{abstract}


\input{1.introduction}

\input{2.relatedwork}
\input{3.background}
\input{4.ordena}
\input{5.visualizationsystem}

\input{6.casestudies}
\input{7.expert_evaluation}
\input{8.discussion}
\input{9.conclusion}

\bibliographystyle{IEEEtran}

\bibliography{references}

\vfill

\end{document}

%% file: 1.introduction.tex


\section{Introduction}
\textcolor{blue}{Understanding how people and goods move through urban environments is a critical task across domains such as transportation planning, logistics, and public safety.} 
Transport problems have become more widespread and severe than ever. The general increase in road traffic and transport demand has resulted in congestion, delays, accidents, and environmental problems beyond what has been considered acceptable. Ortúzar and Willumsen ~\cite{ortuzar2011willumsen} pointed out that (1) economic growth has generated demand levels exceeding most transport facilities' capacity, and (2) long periods of under-investment in some transport modes have resulted in fragile supply systems. Therefore, the role of transport planning is to satisfy certain demands for persons and goods given a transport system with a certain operating capacity. \textcolor{blue}{As urban mobility becomes increasingly complex, new visualization and analysis techniques are needed to better understand the dynamics of movement and their relationship with external factors.}

Origin-Destination movement (or OD flow) data tells a story of geographic displacement, describing demand (origin) and supply (destination). It is usually provided as pairs of geographical locations with associated attributes such as the start and end time of the movement, the trajectory, and conditions~\cite{ferreira2013visual}.   
Identifying and analyzing patterns in OD flows are challenging, and several visualization techniques have been proposed to tackle this problem~\cite{andrienko2013visual,chen2015survey,lu2015trajrank, zeng2017visualizing, lu2016exploring,liu2018tpflow,lu2015od,zeng2019route}.

However, OD flows are not data isolated from other urban conditions such as climate, socioeconomic data, or points of interest. Understanding the relation between external variables and OD flow patterns is an important problem that existing OD visualization tools have not properly tackled. Existing techniques mostly consider the trip data only~\cite{lu2015trajrank}, or the relationship between human mobility and a few external variables such as points of interest (bus stops locations)~\cite{zeng2017visualizing}. 

The large volume of data associated with origin-destination movements brings several analytical challenges, including visual clutter, explainability of observed phenomena, and pattern identification. Despite the advances, existing visualization tools partially address the complexity of the problem but still bear important limitations, mainly when handling OD flows with a large number of records spread in large spatial domains~\cite{zhou2024explainable}. Most techniques resort to data aggregation~\cite{von2015mobilitygraphs} and filtering mechanisms~\cite{guo2009flow} to get around the \textcolor{blue}{visual clutter problem} that naturally appears when visualizing massive OD data. 
On one hand, data aggregation tends to smooth out certain patterns and phenomena, causing misinterpretation of data and a distortion of reality. The spatial partitioning needed to aggregate the flows can disrupt information, especially for short-distance flows. On the other hand, interactive filtering allows for focusing the analysis on particular pieces of data, but it hides the global view of the entire environment. Deciding which pieces should be explored is also not a trivial task. 
Moreover, during OD movements, abnormal events such as crimes or accidents may occur\textcolor{blue}{, and understanding when, where, and
why such events take place is a question of high practical
relevance}. Incorporating such phenomena into the OD flow analytical process is typically of great interest, although only a few works have been devoted to this end~\cite{wu2022enhancing}.

\textcolor{blue}{To address these limitations, we present \ordena\!\!, a visual analytics tool with a dual purpose: (i) to enable intuitive exploration and pattern discovery in OD flows using a novel scatter plot layout, and (ii) to support the analysis of anomalous behaviors, such as traffic accidents or crimes, through integration with classification models and explainable AI (XAI) techniques. While originally motivated by traffic accident analysis, \ordena is general-purpose and applicable to diverse OD-based contexts including crime mapping, logistics failure analysis, and infrastructure planning.}

\ordena is built upon a simple and intuitive OD scatter plot, called \textit{OD-plot}, where the horizontal and vertical axes correspond to origins and destinations, respectively. Therefore, each origin-destination flow is represented as a point in the OD-plot and the way the points are organized in the layout reveals important spatial phenomena present in the data. A sorting mechanism based on the Fiedler vector guarantees that origins and destinations that are spatially close are also placed close to each other on the corresponding axes of the scatter plot, thus enabling the visual identification of patterns such as spatial locations that concentrate short and long-distance OD flows. Besides unveiling distance-related patterns, the proposed OD-plot clarifies which regions concentrate origins whose destinations are spatially scattered and vice-versa. 

\textcolor{blue}{In addition to visual exploration, \ordena allows users to analyze specific behaviors such as traffic accidents or crimes by training classification models that predict the likelihood of such events for each OD flow. These predictions are paired with explanation methods such as SHAP, helping users understand which features contribute most to an anomalous behavior, and how. The system also supports interactive filtering and brushing to focus on selected flows without hiding the full data context.}

\ordena\!\!’s design is based on requirements identified in collaboration with transportation and urban data professionals. \textcolor{blue}{It offers a flexible and interpretable framework that bridges movement data, contextual variables, and analytical reasoning.}

In summary, the main contributions of this work are:
\begin{itemize}
    \item The OD-plot, a scatter plot based tool able to represent OD flows as points in a 2D layout while preserving their spatial neighborhood relationship. Different global and local OD spatial patterns can be identified from the OD-plot layout.
    \item \ordena\!\!, a visual analytic system that combines global and local exploratory mechanisms to scrutinize OD flows and associated attributes.
    \item Integrating visualization, machine learning, and model explainability techniques on a unified platform, joining efforts to analyze abnormal events occurring in OD flows.
    \item Case studies showing the effectiveness of \ordena in exploring and revealing patterns from OD data, \textcolor{blue}{using two datasets with different granularities, spatial units, and incidents (accidents/crime).}
\end{itemize}
\bigskip

%% file: 2.relatedwork.tex
\section{Related Work}
In this section, we discuss techniques designed to specifically visualize origin-destination data, not approaching, and, therefore, more general methods for directed node-link and trajectory visualization. Detailed discussion about directed node-link and trajectory visualization methods can be found in surveys~\cite{andrienko2013visual,beck2017taxonomy,chen2015survey,dodge2008towards,schottler2021visualizing,yoghourdjian2018exploring,wang2017graphs} and comparative studies~\cite{burch2020state,holten2011extended,netzel2014comparative,gu2023classification}.

\smallskip

\myparagraph{Flow Maps} are among the main alternatives for visualizing OD flows. Flow map visualizations build upon straight or curved lines combined with glyph, color, texture, and size to convey the direction and other information associated with a single or a group of OD  instances~\cite{andrienko2016revealing,heredia2024odmeans,koylu2023flowmapper,koylu2017design,stephen2017automated}. Several studies have been carried out to investigate design principles and space for flow maps, providing guidelines from which one can systematically build visualizations that account for the best visual encoding and the information to be kept, attenuated, or added to the visualization~\cite{jenny2018design,tennekes2021design,yang2018origin}.  

Visual clutter is a main issue for flow maps, and several mechanisms such as edge bundling~\cite{cui2008geometry,luo2020clarifying}, overlap avoidance~\cite{jenny2017force}, filtering and brushing~\cite{graser2019untangling}, density kernels~\cite{ibarra2016visualization,guo2014origin}, and data aggregation~\cite{guo2009flow,von2015mobilitygraphs} have been used to alleviate the problem. Tree-like representations depicting the flow departing from a particular location (the tree's root) have also been employed to reduce visual clutter~\cite{buchin2011flow,phan2005flow}.  However, those alternatives tend to aggregate information, thus hiding certain patterns, such as the overall behavior of flows in terms of distances and how much spread or concentrated the flows are over the whole spatial domain.

%

\smallskip

\myparagraph{OD matrix} based representations rely on the flow network to build a matrix where rows and columns correspond to origin-destination locations, and each matrix entry accounts for the amount of movement from an origin to a destination. Clustering and rows/columns reordering are usually employed to reveal patterns~\cite{luo2017constructing}, while colored heatmaps are often used to encode particular attributes of the OD flow~\cite{perez2020modalflow}.  
The main issue with the OD matrix is the lack of a mapping from the matrix to geographical locations. Alternatives to alleviate the problem include the organization of OD matrices in a geographically structured grid decomposition~\cite{wood2010visualisation}, relying on a straight-line based labeling scheme to connect matrix rows (or columns) to their corresponding geographical locations on the map~\cite{yang2016many}.
There are also matrix-based representations where each row corresponds to an origin-destination pair while columns encode an attribute associated with each OD flow, as the quantity of flow that takes place over time~\cite{boyandin2011flowstrates}.

OD matrix-based visualizations scale better than flow maps and can unveil certain global patterns when properly sorted rows and columns. However, identifying patterns from large fine-grained (non-aggregated) OD data is not straightforward.

\smallskip

\myparagraph{Visual querying} comprises methods designed to assist the analysis of large movement data, enabling the identification of traffic patterns~\cite{ding2017}, route diversity~\cite{liu2011visual}, and traffic jams~\cite{wang2013visual}. Interactive querying is the main mechanism employed to select and explore subsets of movements~\cite{kruger2013trajectorylenses,zeng2016visualizing}, making possible the comparison of groups of OD movements~\cite{ferreira2013visual} and the evaluation of traffic patterns in specific locations~\cite{zeng2013visualizing,wang2014visual}. 
Some works rely on embedding schemes to represent OD data in a latent space, relying on topic modeling to extract spatio-temporal patterns~\cite{shi2019exploring} or
projecting the embedded data into a 2D visual space in order to interactively filter data of interest~\cite{zhou2018visual}. Data filtering through interactive querying enables the identification of patterns contained in specific chunks of data, but it hampers the visualization of global phenomena, mainly spatial ones. 

Our methodology \textcolor{blue}{leverages these three resources, combining them by integrating a flow map with a scatterplot analogous to the matrix but independent of its resolution,} also incorporates \textcolor{blue}{interactive visual querying through} filtering mechanisms, which, combined with explainability resources, enable the identification of particular phenomena.However, two main aspects differ our approach from the existing ones. First, the \ordena system provides a visual resource (the OD-plot) that allows a global analysis of OD flows without data aggregation, making the visual identification of origins (destinations) whose destinations (origins) are scattered in the spatial domain while revealing where short and long trips are more concentrated. Moreover, ORDENA enables interactive resources to explore a large number of attributes as well as phenomena associated with OD flows.

%% file: 3.background.tex
\section{Background}
\label{sec:background}

We outlined requirements crucial for a productive OD flow analytical process from meetings with a transportation engineering expert. During the meetings, we identified external information that should complement the analysis. Based on these insights, we formulated analytical tasks that must be supported by the visual analytic tool, detailed in the following.

\subsection{Dataset and Inputs}
\label{sec:back_dataset}

\textcolor{blue}{ORDENA has been evaluated on two real-world, large-scale datasets that differ in granularity, spatial unit, and problem domain. This allows us to demonstrate both scalability and adaptability.}

\textcolor{blue}{\textbf{Dataset A}} corresponds to traffic accidents in Melbourne, Australia, provided by a collaborating transportation expert. It contains approximately 120,000 instances from 2014 to 2018 \textcolor{blue}{aggregated across 302 census sectors at ZIP-code granularity,} each representing an individual involved in a traffic accident. Origins correspond to the victim’s residential ZIP code and destinations to the ZIP code of the accident site. \textcolor{blue}{Each record is enriched with 78 attributes covering accident details, personal characteristics, and census-related context of both origin and destination.} A detailed attribute list is provided in the \textit{Supplementary Material}.

\textcolor{blue}{\textbf{Dataset B} focuses on bicycle thefts in São Paulo. It comprises 698 trips, each described by origin and destination coordinates at the street-intersection level (119748 intersections mapped at latitude–longitude precision). For each trip, we estimate the likelihood of traversing a path where a theft occurred (a density value between 0 and 1). While this does not guarantee that the cyclist was the actual victim, it captures exposure to theft-prone areas. Each instance includes 147 attributes, such as bicycle cost, departure time, duration, and parking type.}

\textcolor{blue}{For both datasets, ORDENA assumes as input a classification and an explanation model. In Dataset A, the task is to predict whether an accident victim is fatal; in Dataset B, whether a trip is exposed to theft.} After testing different classifiers, we selected a Random Forest with Random Undersampling for its superior performance (see \textit{Supplementary Material}). For interpretability, we employ SHAP~\cite{lundberg2017unified}, chosen for its robustness and flexibility across classifiers, to quantify local feature importance in both domains.

\subsection{\textcolor{blue}{Requirements and} Analytical Tasks}
\label{sec:requirements}

The domain expert involved in this project is primarily concerned with uncovering and analyzing potential patterns in traffic accident fatalities. To create a useful and effective visual analytic tool, we have identified a set of \textcolor{blue}{requirements and corresponding} tasks that the visualization tool must support.
\textcolor{blue}{
\textbf{Requirements:}
\smallskip
\myparagraph{R1. Support spatial exploration of trips and incidents.} 
Enable filtering of trips and identification of global and local origin-destination patterns.
\smallskip
\myparagraph{R2. Enable detailed analysis of trip attributes and incident factors.} 
Allow exploration of attributes and comparison of trips with or without incidents.
\smallskip
\myparagraph{R3. Facilitate model evaluation and classification analysis.} 
Support the analysis of classification performance across subsets and spatial locations.
}

\smallskip
\textbf{Tasks:}
\smallskip
\myparagraph{T1. Interactive Trip Filtering.} 
Select subsets of trips and visualize their spatial location. Enable a mechanism to show all trips or only those related to a particular type of event (fatality \textcolor{blue}{or theft} in our context). \textcolor{blue}{Implements R1.}

\smallskip

\myparagraph{T2. Identify Global and Local Origin-Destination Patterns.}
Identify global patterns in trip data, highlighting regions with a high concentration of short trips and locations where long trips either originate or terminate. Pinpoint origins that have widely dispersed destinations, as well as destinations that attract trips from various origins. \textcolor{blue}{Addresses R1.}

\smallskip

\myparagraph{T3. Exploration of Trip Attributes.}
Given a subset of trips, identify their most relevant attributes. Given a particular attribute, visualize its statistical distribution. \textcolor{blue}{Fulfills R2.}

\smallskip

\myparagraph{T4. Compare \textcolor{blue}{trips} with or without \textcolor{blue}{incidents}}
Identify spatial patterns of origin and destination where \textcolor{blue}{abnormal behavior occurs}. Identify the most relevant factors (attributes) characterizing fatal car accidents \textcolor{blue}{or trips with bicycle theft} and compare these factors with those of non-fatal accidents \textcolor{blue}{or trips without theft occurring} in similar origin and destination contexts. \textcolor{blue}{Supports R2.}

\smallskip

\myparagraph{T5. Classify and Analyze Fatalities.}
Analyze the performance of the classification model in terms of failures and successes, verifying whether the performance varies over different subsets and spatial locations. \textcolor{blue}{Meets R3.}

\subsection{System Overview}
\label{subsec:system_overview}
\ordena\! is a visualization-assisted analytical system designed to address the abovementioned tasks. It is built upon two main concepts: an origin-destination scatter plot (OD-Plot) and a classification/explanation mechanism for \textcolor{blue}{incidents or events of interest}.

\textcolor{blue}{
The novelty of our approach lies in the integration of multiple resources: visualization, classification, and explainability, resulting in more accurate and context-aware behavioral explanations. 
Visualization reveals mobility patterns, classification categorizes accidents or crimes, and explainability interprets the features contributing to the results. 
Our goal is to provide actionable insights that support decision-making, rather than merely presenting categorical outcomes.}

The OD-plot is a scatter plot where each depicted point represents a trip (see Fig.~\ref{fig:visualization} (a)). The horizontal and vertical axes are structured such that points close to each other in the plot correspond to spatially similar trips, that is, trips starting from nearby origins and arriving at nearby destinations. From the OD-Plot one can visually identify trip patterns and interactively filter subsets of trips. All visualization components (see Fig.~\ref{fig:visualization}) are updated from a selection made on the OD-plot. The OD-Plot has been designed to make the tasks T1, T2, and part of T4 feasible.

\ordena also relies on a combination of classification and explanation mechanisms to enable the analysis of the attributes related to \textcolor{blue}{incidents}. In other words, the explanation scheme is applied to a machine learning model that seeks to classify the accidents as fatal or non-fatal \textcolor{blue}{or as involving theft or not}. The results and corresponding explanations are depicted in interactive visual components showing the model's accuracy, feature importance, and statistical distributions. Therefore, the classification model and corresponding explanation tackle the tasks T3, T4, and T5. 

\ordena\!\!'s \textcolor{blue}{visualization} workflow \textcolor{blue}{is highlighted in yellow in Fig.~\ref{fig:visualization}. The linked interaction comprises five visual components (gray points A, B, C, D, E) and four levels of depth (yellow points 0, 1, 2, 3), representing the sequence of interaction steps. Level 0 involves selecting trips via checkboxes, with or without incidents, which updates the OD-Plot to show only the selected group. Level 1} consists of the user selecting a subset of trips of interest from the OD-plot \textcolor{blue}{(A)}. The selection is reflected in all other components (see \secref{VisualizationSystem}). \textcolor{blue}{At Level 2, three possible branches are triggered: the first shows the spatial distribution of the selected trips (B), the second presents the most relevant features (C), and the third displays the classification model's performance on the selected subset (D). Level 3 is triggered by clicking a bar in (C), displaying the attribute’s statistical distribution in (E).}

Before describing \ordena\!\!`s visual components, we further detail the mathematical and computational foundations behind the OD-plot, the classification model, and the explanation mechanism that supports the proposed visualization tool.


%% file: 4.ordena.tex
\section{OD-Plot and Trips Explanation}
\label{sec:od-plot}
 As mentioned in Section~\ref{sec:background}, \ordena is built upon two main concepts, the OD-plot, and trip classification and explanation, which we detail in the following.

\subsection{OD-Plot}

Each origin-destination data instance can be seen as a single tuple point $(O,D)$ with attributes associated with it. Tuples are naturally depicted in scatter plots where the Cartesian axes account for the coordinates of the tuple. In our context, the abscissa corresponds to the origins and the ordinate to the destination. Therefore, each trip is represented by a \textcolor{blue}{single two-dimensional point} in the OD-Plot.
The main challenge when building the OD-plot is that origins and destinations are each given by a pair of spatial coordinates, typically a latitude/longitude pair. Thus, associating each origin and destination to single coordinates on the axes demands some mapping mechanism.

In our context, origins and destinations are given on the nodes of a graph, \textcolor{blue}{corresponding to city sectors in the Melbourne dataset (A) or to street intersections in the São Paulo dataset (B). They }are uniquely represented by the node identifier number (node ID). 

Two conditions are needed to guarantee that trips with nearby origins and destinations will be depicted close to each other in the OD-plot. First, the graph nodes must be enumerated such that adjacent nodes have nearly consecutive IDs. Second, the OD-plot axes must be discretized following such IDs' order. The problem is how to enumerate the graph nodes while respecting their neighborhood relation.

Fortunately, the solution to this problem is well known. It derives from an eigenvector of the graph Laplacian matrix, the so-called Fiedler vector. The Fiedler vector has been employed in a number of visualization applications, for example, to semantically order words in a world cloud~\cite{paulovich2012semantic},  arrange rows and columns in a matrix-based network visualization~\cite{behrisch2016matrix}, and flatten 3D structures into 2D visual spaces~\cite{kreiser2018survey}.

In mathematical terms, the Fiedler vector can be defined as follows: let $\mathcal{G}$ be a connected graph and $\mathbf{L}$ be the Laplacian matrix associated with $\mathcal{G}$. In our context, $\mathcal{G}$ corresponds to the graph generated from the spatial discretization given in the dataset, where the nodes correspond to zip code tracts, with edges connecting tracts that intersect each other \textcolor{blue}{in Dataset A. In Dataset B, the graph has a finer granularity, with nodes representing street corners and edges representing to the streets connecting them.} Edge weights corresponding to non-zero off-diagonal elements of $\mathbf{L}$ are given by $L_{ij}=-1/l_{ij}$, where $l_{ij}$ is the length of the edge connecting the nodes $i$ and $j$. If the $i$ and $j$ nodes are not connected, then $L_{ij}=0$. Diagonal elements are settled as $L_{ii}=\sum_{j\neq i} |L_{ij}|$. The Laplacian matrix $\mathbf{L}$ is positive semi-definite, so its eigenvalues are real and non-negative. If  $\mathcal{G}$ is connected, then $\mathbf{L}$ has exactly one eigenvalue equal to zero. The Fiedler vector corresponds to the eigenvector of $\mathbf{L}$ that is associated to the smallest non-zero eigenvalue. A detailed discussion about Laplacian matrices and their spectrum, including the Fiedler vector, can be found in Chung's book~\cite{chung1997spectral}.

\begin{figure}[!t]
    \centering
    \includegraphics[width=\linewidth]{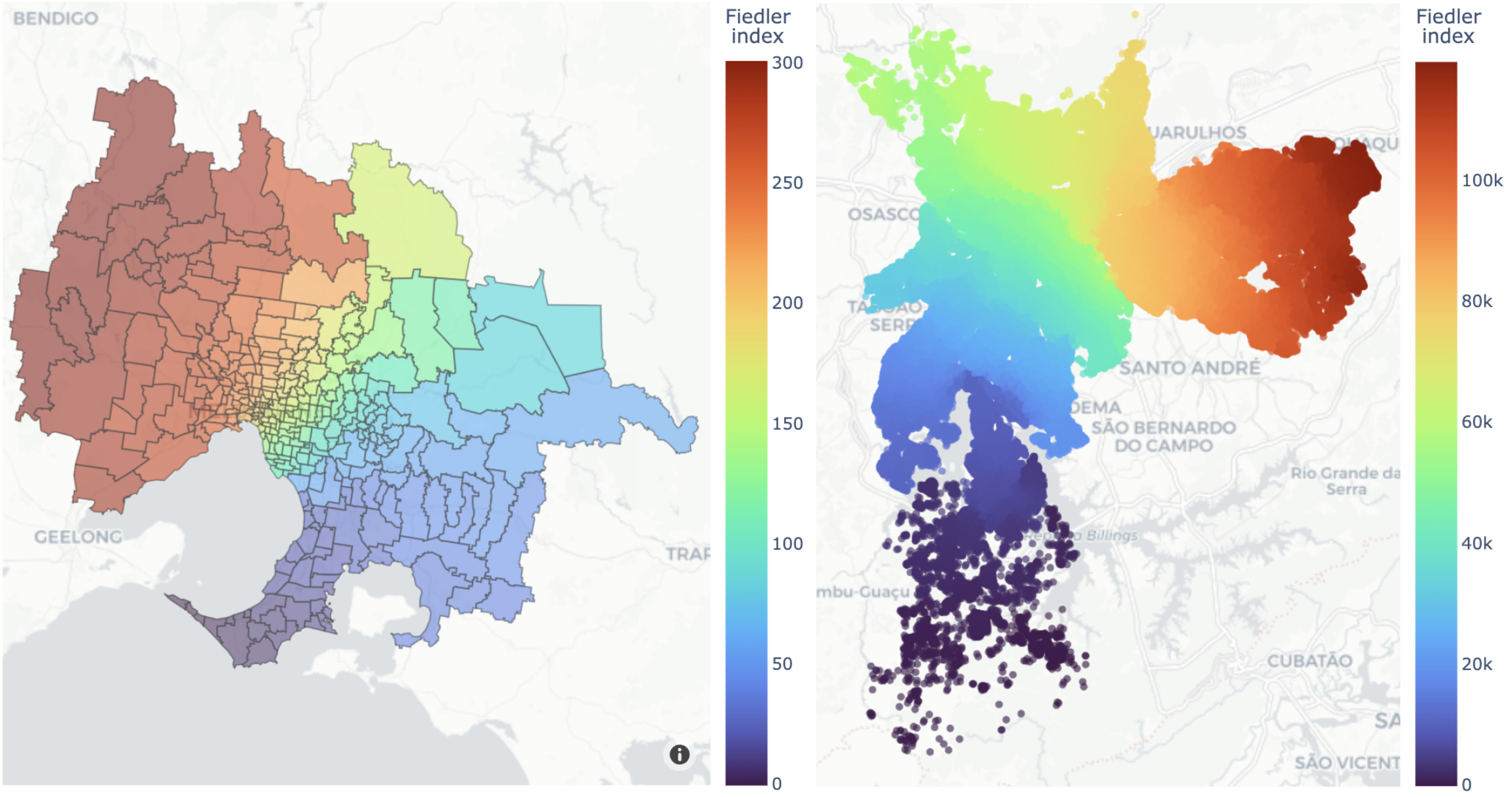}
    \caption{
    \textcolor{blue}{Visualization of spatial units characterized by the Fiedler vector (zip codes or street intersections). On the left, zip code tracts from the Melbourne accidents dataset; on the right, street intersections from the São Paulo bicycle theft dataset. In both maps, units are color-coded according to their Fiedler vector indices, with nearby units showing similar colors and the chromatic scales displayed using a continuous colormap.}
    }
    \label{fig:fiedler}
\end{figure}

Denoting the Fiedler vector by $\mathbf{v}=[v_1,\ldots, v_n]$, where $v_i$ is the value associated to noted $i$ (each entry in $\mathbf{v}$ is associated to a node in $\mathcal{G}$) and $n$ is the number of nodes in $\mathcal{G}$, it can be shown that:
\begin{equation}
\mathbf{v} = \min_{\|u\|=1}\,\,\sum_{\substack{\forall\, i\neq j \mbox{ s.t.}\\ L_{ij}\neq 0}} |L_{ij}|(u_i - u_j)^2 \, .
\label{eq:fiedler}
\end{equation}
Eq.~(\ref{eq:fiedler}) says that the values $v_i$ and $v_j$ are pushed to be similar if the nodes $i$ and $j$ are linked by an edge in $\mathcal{G}$. Moreover, the closer they are spatially (higher $|L_{ij}|$ values), the more similar $v_i$ and $v_j$ will be. Sorting the entries in $\mathbf{v}$ and numbering the nodes according to the sorted order, one makes neighbor nodes labeled with nearly consecutive IDs. Discretizing the OD-plot axes in ascending (or descending) ID order ensures that trips with nearby origins and destinations will be represented by close points on the plot. Figure~\ref{fig:fiedler} illustrates the results \textcolor{blue}{at two different levels of spatial granularity, demonstrating the versatility of the Fiedler vector.} Importantly, we exploit transparency to provide density information and avoid clutter visually (see Section \ref{subsec:ODplot}).

\textcolor{blue}{Spatial graph vertex ordering algorithms, which map two-dimensional spatial coordinates into a single one-dimensional index, have been explored through both topological approaches (e.g., Fiedler ordering) and geometric approaches (e.g., 1D t-SNE and UMAP). The evaluation of how well these indices preserve spatial proximity, that is, whether nodes that are close in 2D remain close in the 1D ordering, was investigated in prior work \cite{salinas2024visual} and falls outside the scope of the present paper. That previous work established the relationship between geographic distances and index distances through both quantitative and qualitative (visual) metrics.}

\textcolor{blue}{Building on the demonstrated effectiveness of Fiedler ordering (see \cite{salinas2024visual}), and recognizing the necessity of preserving topological properties, we adopt Fiedler for mapping coordinates in the OD-plot. Unlike purely geometric methods, Fiedler leverages the spectral properties of the graph Laplacian to capture global structural relationships within the mobility network. This topological perspective is essential, as it enables the preservation of structural roles (such as hubs, transfer points, and peripheral nodes) which are crucial for maintaining meaningful connectivity patterns beyond mere spatial proximity. Thus, the OD-plot effectively reflects this topological ordering, serving as a visual proxy for the underlying spatial relationships and motivating our choice of this representation.} 

\smallskip

\myparagraph{OD-Plot Interpretation and Expressiveness}

\begin{figure}[t]
    \centering  \includegraphics[width=0.6\columnwidth]{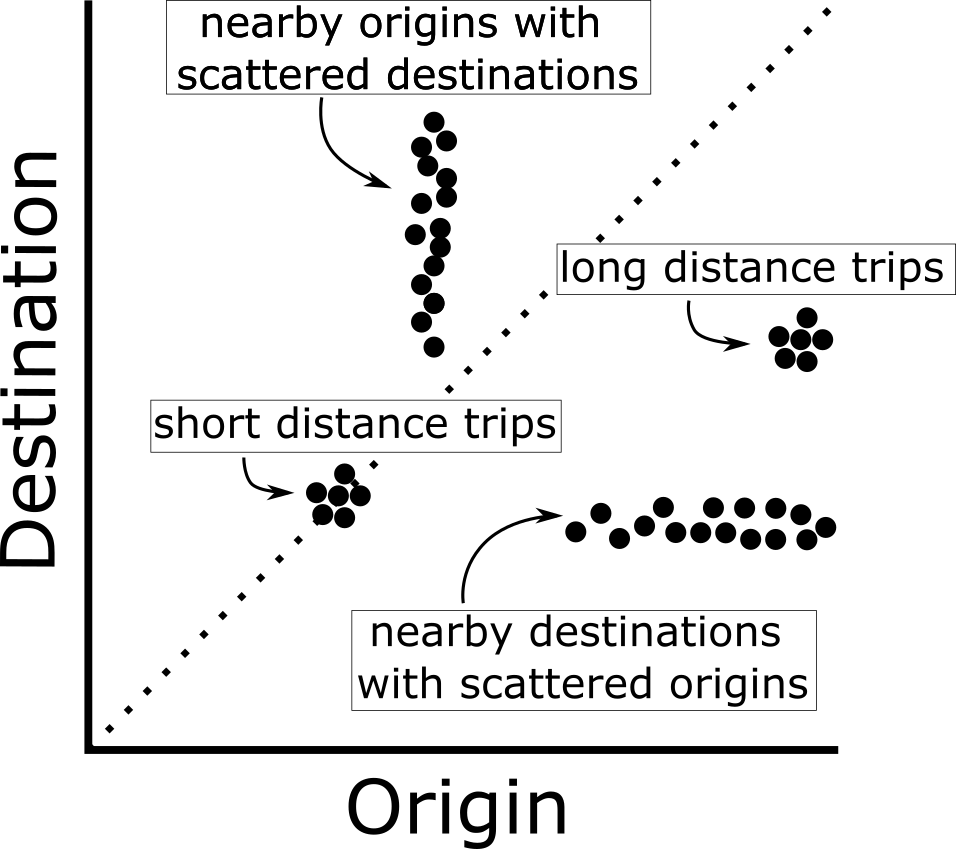}
    \caption{OD-Plot interpretation and expressiveness. The arrangement of points corresponding to OD flows on the plot reveals different spatial patterns.}
    \label{fig:od-plot}
    \vspace{-0.7cm}
\end{figure}

The OD-plot provides a global view of OD flows, enabling the visual identification of different patterns. 
Examples of patterns that are naturally unveiled on the OD-plot are illustrated in Fig.~\ref{fig:od-plot}. For instance, a set of points aligned vertically corresponds to trips with nearby origins whose destinations are scattered in the spatial domain. Points aligned horizontally correspond to the opposite, that is, trips with nearby destinations whose origins are spread out. Points concentrated around the diagonal represent short-distance trips, while groups of points away from the diagonal account for longer trips leaving from and arriving in particular regions. \textcolor{blue}{The OD-plot organizes trips not only by length (distinguishing short versus long-distance travel) but also by spatial proximity (geographically close OD pairs tend to remain close in the OD-plot). It reveals hierarchical mobility structures that capture local and global flows, supporting the identification of nested or overlapping patterns. By grouping trips according to their structural similarity, the OD-plot facilitates the detection of behavioral or contextual patterns that may be associated with risk factors, for instance, trips toward peripheral areas may carry higher risk, or very short trips may indicate suspicious activity. Moreover, while the OD-plot abstracts away explicit geographic information, our complementary Trip Location View maps OD trips directly onto geographic space, preserving spatial fidelity and reducing cognitive effort. Together, these views enable faster, multi-scale analysis of mobility patterns, uncover latent hierarchies, and provide advantages over traditional map-based or purely spatial filtering approaches, revealing structures in OD data that support reasoning about accidents, crimes, or other anomalies.} Other patterns also appear on the OD-plot, which will be discussed in the case studies (\secref{cases}).

\begin{figure}
    \centering
    \includegraphics[width=0.32\linewidth]{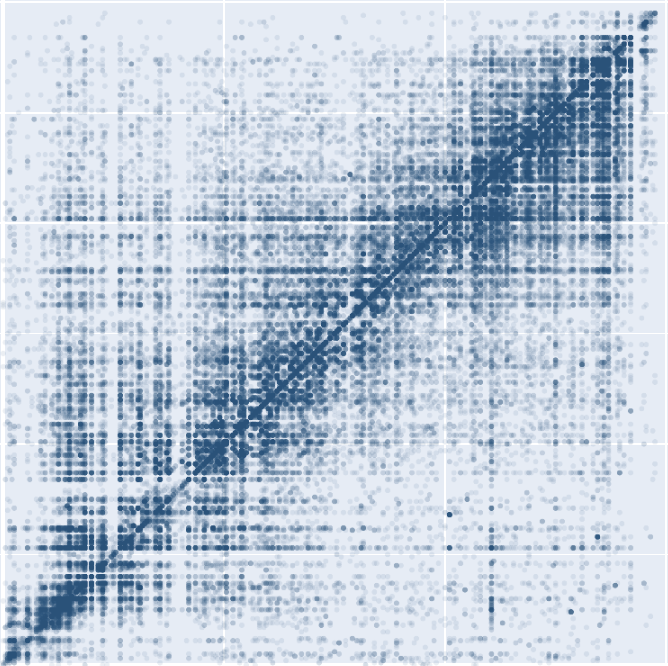}
    \includegraphics[width=0.32\linewidth]{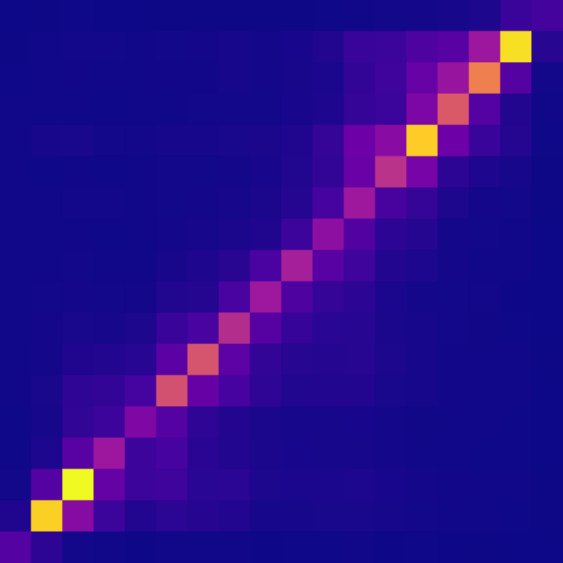}
    \includegraphics[width=0.32\linewidth]{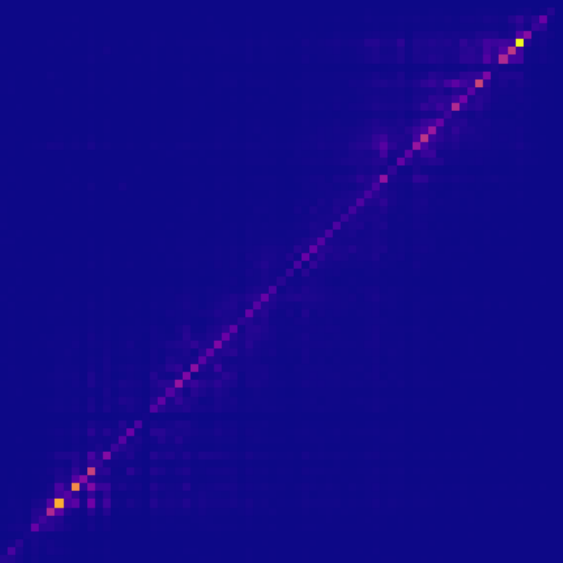}
    \caption{OD-plot (left) and matrix representations with resolutions 18$\times$18 (middle) and 71$\times$71 (right). OD patterns are more clearly revealed on the OD-plot.}
    \label{fig:comparison}
    \vspace{-0.7cm}
\end{figure}

The Fiedler-based scheme can also be employed to sort the rows (destinations) and columns (origins) of OD flows matrix representation. In contrast to the proposed OD-plot, the matrix representation requires a resolution to be specified beforehand. Defining an appropriate resolution level is not easy, as a low resolution aggregates the information, thus smoothing out certain patterns, while a high resolution demands an additional effort to tune color maps to properly visualize trip patterns. Fig.~\ref{fig:comparison} illustrates this issue, where an OD-plot and two matrix representations generated from the same dataset are depicted. The OD-plot on the left is discretized as 302$\times$302 positions (302 is the number of zip code tracts \textcolor{blue}{from Dataset A}), while the matrix representations in the middle and right have a resolution of 18$\times$18 and 71$\times$71 cells, respectively.

Notice that, in both matrices, the diagonal is the only clear visible pattern, representing a high concentration of short-distance trips. This phenomenon can be attributed to the default color mapping strategy, which linearly associates cell densities with colors. In contrast, the OD-plot excels in uncovering additional meaningful patterns, such as trips from a wide variety of places whose destinations are the same and an absence of trips from specific origins. Furthermore, it is readily apparent that the higher-resolution matrix erroneously creates the illusion of  ``gaps'' along the diagonal when compared to the lower-resolution matrix.

\subsection{Classification and Explanation}
 


Exploring attributes related to Origin-Destination (OD) data is crucial for analyzing and understanding patterns and observed phenomena. Take, for instance, the OD traffic accident \textcolor{blue}{Dataset A}, where a key concern is discerning the attributes associated with fatal and non-fatal incidents, \textcolor{blue}{or the OD bike \mbox{Dataset B}, where the focus lies on identifying features linked to theft and non-theft cases}. With \ordena, users can investigate the importance of features within interactively selected subsets of data on the OD-plot. This capability empowers users to gain insights into the behavior of attributes tied to specific spatial patterns.

Applying techniques that explain classification models is a natural choice to accomplish such a task. To enable this exploration, we induced a suitable classification model, as discussed in \secref{back_dataset} and detailed in the \textit{Supplementary Material}. Subsequently, we apply the SHAP explanation method~\cite{lundberg2017unified} for extracting the feature importance values. 
Shapley values are computed for each instance, and the values for a subset of data requiring explanation are averaged. Acknowledging that Shapley values can be negative, the averaging process employs the absolute values. Notably, the average is computed separately for each class, enabling a direct comparison of feature importance between fatal and non-fatal accidents.


%% file: 5.visualizationsystem.tex
\section{\ordena\!\!'s Visual components}
\label{sec:VisualizationSystem}

\begin{figure*}[!t]
    \centering
    \includegraphics[width=1.5\columnwidth]{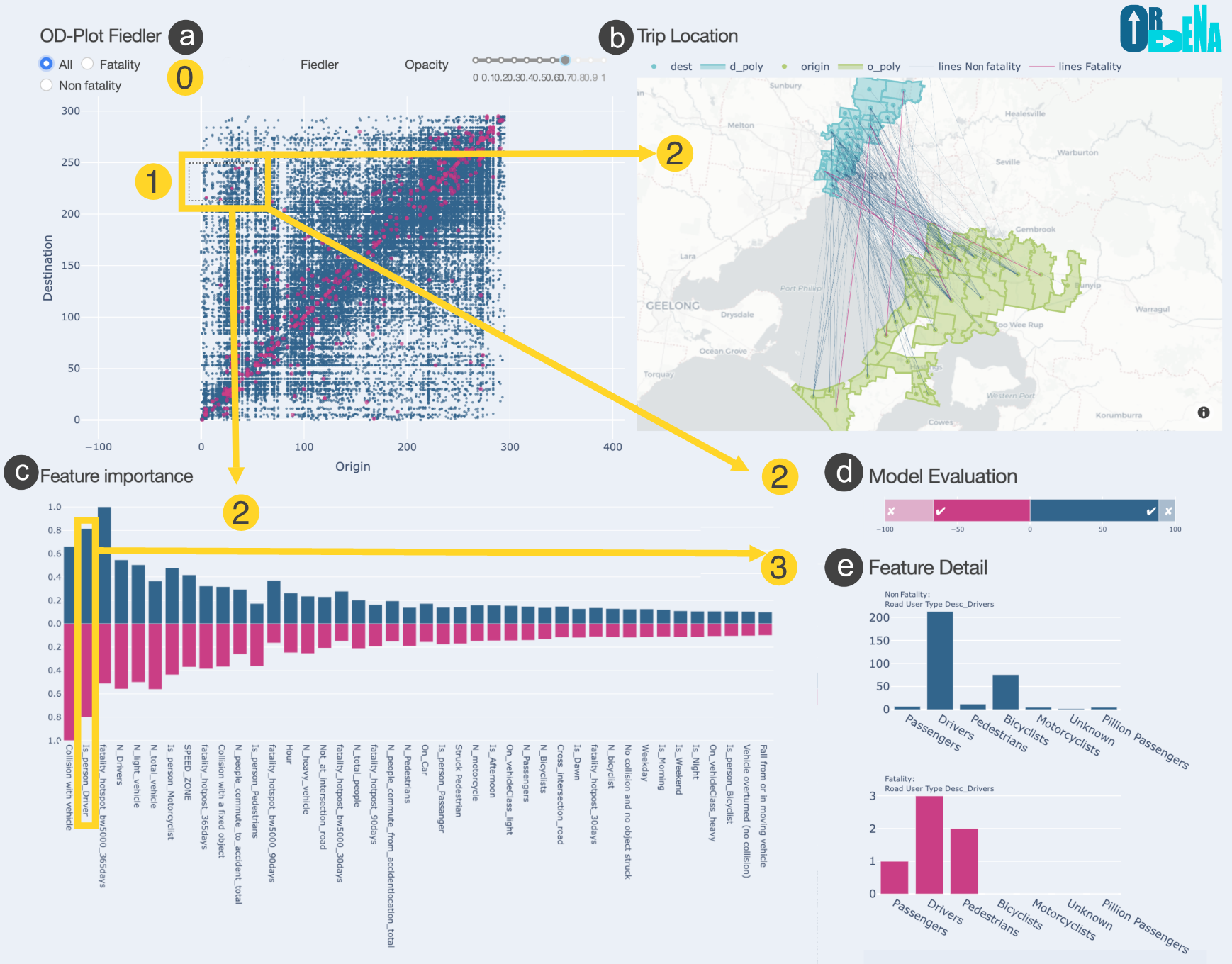}
    \caption{\textcolor{blue}{Components of the \ordena system: (a) OD-Plot, (b) Trip Location, (c) Feature Importance, (d) Model Evaluation, and (e) Feature Detail views. The yellow-highlighted workflow illustrates four interaction levels, guiding the user through the exploration of OD flows.}}
    \label{fig:visualization}
\end{figure*}

The \ordena system, depicted in the Figure~\ref{fig:visualization}, contains five linked visual components that integrate trips and feature exploration: (a) OD-Plot View, to visualize trip patterns according to their origin and destination; (b) Trip Location View, to display trips on the map; (c) Feature Importance View, to reveal the most important features of the selected set of trips; (d) Model Evaluation View, to assess the performance of the predictive model through the percentage of hits and misses in each class; and (e) Feature Detail View, describing the distribution of a specific feature over a selected subset of trips. These visual components have been designed to properly address the \textcolor{blue}{requirements and ensuing tasks} raised from the interaction with the domain experts, as discussed in \secref{requirements}. Table~\ref{tab:taskvscomponent} indicates the relation between the visual components (rows) and the tasks (columns), \textcolor{blue}{with the requirements listed at the bottom of each column}. In what follows, we detail each component.

\begin{table}[!t]
\begin{tabular}{llccccc}
\hline
 
                                 & \textbf{Sec.}                & \textbf{T1} & \textbf{T2} & \textbf{T3} & \textbf{T4} & \textbf{T5} \\ \hline
\textbf{OD-Plot View}            & \ref{subsec:ODplot}          & \checkmark  & \checkmark  &             &      \checkmark       &             \\
\textbf{Trip Location View}      & \ref{subsec:TripLocation}    &  \checkmark     & \checkmark  &             &      \checkmark       &             \\
\textbf{Feature Importance View} & \ref{subsec:FeatureImportance}&             &             & \checkmark  & \checkmark  &             \\
\textbf{Model Evaluation View}   & \ref{subsec:ModelEvaluation} &             &             &             &             & \checkmark  \\
\textbf{Feature Detail View}     & \ref{subsec:FeatureDetailComponent}&             &             & \checkmark  & \checkmark  &             \\ \hline
\textcolor{blue}{\textbf{Obeys Requirement}} & \textcolor{blue}{-} & \textcolor{blue}{R1} & \textcolor{blue}{R1} & \textcolor{blue}{R2} & \textcolor{blue}{R2} & \textcolor{blue}{R3} \\ \hline
\end{tabular}
\caption{\label{tab:taskvscomponent}\textcolor{blue}{Requirements,} analytical tasks and visual components.}
\vglue -0.5cm
\end{table}

\subsection{OD-Plot View} 
\label{subsec:ODplot}
This view, shown in Figure~\ref{fig:visualization}(a), allows visualization of travel patterns from a scatter plot. As discussed in Section~\ref{sec:od-plot}, each \textcolor{blue}{trip} corresponds to a point in the scatter plot whose axes represent its origin (abscissa) and destination (ordinate), respectively. The position in each axis is derived from the Fiedler vector computed from the street graph (see details in Section~\ref{sec:od-plot}). Interaction resources such as zoom, pan, box select, lasso select, among others, can be used to explore the OD-Plot and to select trips in delimited areas or by axes. Additionally, the upper part of this view contains radio buttons to select whether to show all trips, trips \textcolor{blue}{with incidents} (purple), or trips \textcolor{blue}{without incidents} (blue). 

A slider bar controls the level of opacity of the points, allowing better visualization of regions with a dense concentration of trips in the plot, \textcolor{blue}{such as the diagonal bands, which can arise both from genuine short-distance mobility patterns and from artifacts of the spectral ordering. Adjusting opacity mitigates overplotting and helps reveal secondary structures within dense areas.} Once a set of trips has been selected, the other components are activated or updated.

\subsection{Trip Location View}
\label{subsec:TripLocation}
Illustrated in Figure~\ref{fig:visualization}(b), the Trip Location View presents the geographical distribution of \textcolor{blue}{the selected trips}. The \textcolor{blue}{origin locations (polygons for zip code granularity and points for latitude-longitude granularity)} are shown in green, the \textcolor{blue}{destination locations} in cyan, and the connecting lines are colored according to the trip class: \textcolor{blue}{for Dataset A, fatal accidents (purple) versus non-fatal trips (blue), and for Dataset B, theft-related trips (purple) versus regular trips (blue)}. Users can interact with the legend at the top to show or hide specific markers. The trips displayed in this view correspond to those selected in the OD-Plot View, following a cross-filtering approach. This view is particularly effective for examining the geographic distribution of a specific subset of trips.

\subsection{Feature importance View}
\label{subsec:FeatureImportance}
The Feature Importance View, depicted in Figure~\ref{fig:visualization}(c), provides explainability information from the predictive model. Given a set of trips selected on the OD-Plot View, trips' attributes are ranked by the explanation method (see details in Section~\ref{sec:od-plot}) and the ranks sorted in decreasing order are shown in the view. 
The blue bars correspond to the feature importance of trips without \textcolor{blue}{incident} while the purple bars account for trips with \textcolor{blue}{incident}.
Therefore, the ordered features reveal which factors contributed most to the predictive model. By clicking on a specific bar, the statistical distribution of the corresponding attribute is displayed in the Feature Detail View (discussed in Sec.~\ref{subsec:FeatureDetailComponent}).

\subsection{Model Evaluation View}
\label{subsec:ModelEvaluation}
This component, shown in Figure~\ref{fig:visualization}(d), outlines the percentage of successes and failures of the classification model over a subset of trips selected on the OD-Plot View. The percentage of successes (\checkmark), that is, when the true class coincides with the predicted one, are encoded in the opaque part of the bar, and the errors ($\times$) are shown in the partially transparent part. The color corresponds to the trip class. 

\subsection{Feature Detail View}
\label{subsec:FeatureDetailComponent}
This view, depicted in Figure~\ref{fig:visualization}(e), allows delving one level further into the explanation, showing the statistical distribution of a selected attribute for both trips \textcolor{blue}{with and without incidents}. Attributes with discrete values are displayed as frequency bar charts, while continuous-valued attributes are shown as regular histograms. For example, \textcolor{blue}{in Figure~\ref{fig:visualization}}, by clicking on the \textit{Is\_person\_Driver} bar in the Feature Importance View, the resulting Feature Detail View displays two charts, one for each trip type. \textcolor{blue}{Note that \textit{Is\_person\_Driver} itself does not contain any statistics; it represents a question about whether the affected person was the driver. The Feature Detail View then visualizes the distribution of deceased individuals according to their role; Passengers, Drivers, Pedestrians, Bicyclists, Motorcyclists, Unknown, and Pillion Passengers, for each trip type.} Both charts share the same horizontal axis labels for consistency, and the number of trips for each attribute is shown on the vertical axis.

\section{Implementation Details}
\ordena is a web-based system implemented using Python and JavaScript. The system is composed of two main components: data modeling and visualization.

For the data modeling component, \textcolor{blue}{Dataset A employs city-region postcodes as the spatial unit for aggregation, whereas Dataset B relies on street corners as spatial references.} We further incorporate the feature vector of each trip together with their SHAP values. This pre-processing step generates a structured file that serves as input to the visualization component.

The visualization has been developed in Python, employing the PlotlyJs (plotly.com/javascript) and PlotlyDash (plotly.com/dash) libraries within the Jupyter Notebook platform. The Dash Bootstrap Components library is used to design the interface and its integration. The Trip Location View is supported by the OpenStreetMap database. For interaction tasks in the OD-Plot and Trip Location Views, the system leverages the Plotly Modebar.

\textcolor{blue}{To ensure seamless interaction, \ordena adopts a hybrid computational strategy in which classification and ordering results are precomputed, while user-driven selections and filters are processed in real time. This architecture mitigates computational overhead in the visualization pipeline, ensuring that all interactions (e.g., filtering, brushing, cross-component updates) remain responsive. Performance evaluations demonstrate sub-second response times across tasks, including mobile access, as \ordena is deployed on a web server and is therefore independent of client-side hardware. Consequently, \ordena delivers interactive performance on large-scale OD datasets while maintaining accessibility and usability.}

%% file: 6.casestudies.tex
\section{Case Studies}
\label{sec:cases}

This section presents and discusses three case studies to showcase how \ordena enables exploratory and analytical resources to tackle the tasks detailed in \secref{requirements}. 
First, we show an example of exploratory analysis to find spatial patterns, relying mainly on the OD-Plot and Trip Location View. 
The second case study aims at analyzing and extracting patterns present in trips with \textcolor{blue}{incidents}. In the third case study, we illustrate how \ordena can be employed to evaluate the predictive model in subsets of data. In the case studies, we explicitly point out the tasks enumerated in \secref{requirements}, making clear how \ordena is addressing each one.

\subsection{Exploring Spatial Patterns of OD Flows}
\label{subsec:caseA}
This case study aims to visually identify and analyze both global and local spatial patterns of OD flows, relying mainly on the OD-Plot View and the Trip Location View.

\begin{figure}[!t]
    \centering
    \includegraphics[width=0.9\columnwidth]{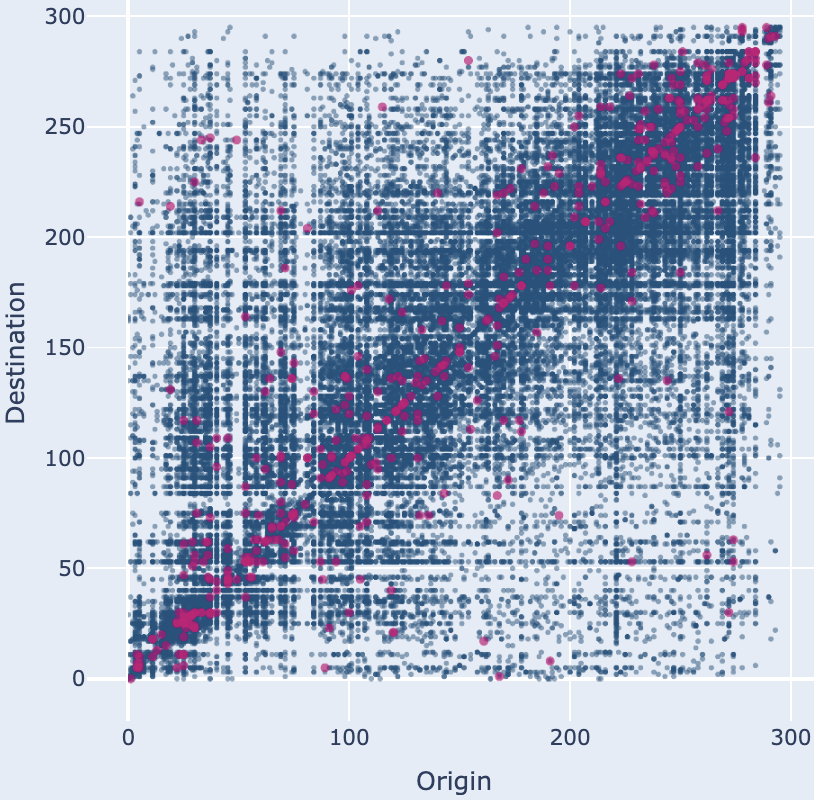}
    \caption{\textcolor{blue}{Case Study \ref{subsec:caseA} (A): OD-Plot visualization of car trips in Melbourne (Dataset A). Regular trips are represented by blue dots, while trips with fatalities are depicted in purple. Diagonal concentration indicates predominance of short-distance trips.}}
    \vspace{-0.7cm}
    \label{fig:casestudy1_odplot}
\end{figure}
\textcolor{blue}{\textbf{Dataset A.}} \figref{casestudy1_odplot} shows the OD-Plot derived from the Melbourne traffic accident (\secref{back_dataset}), including both accidents without fatalities (blue), which represent the majority of trips, and accidents with fatalities (purple).  

The most prominent global pattern relates to trip distance: most trips are located near the diagonal, indicating a predominance of short-distance trips. \textcolor{blue}{Nevertheless, the OD-Plot also reveals a significant number of accidents in long-distance trips, corresponding to points located far from the diagonal.} Another relevant global pattern is the asymmetry with respect to the diagonal. The upper triangle above the diagonal is noticeably denser than the lower triangle, indicating that many accidents on long-distance trips tend to occur in a specific direction. \textcolor{blue}{For instance, trips with destinations near index 200 register more accidents than trips originating from this region.}  

\begin{figure}[!t]
    \centering
    \includegraphics[width=1.0\columnwidth]{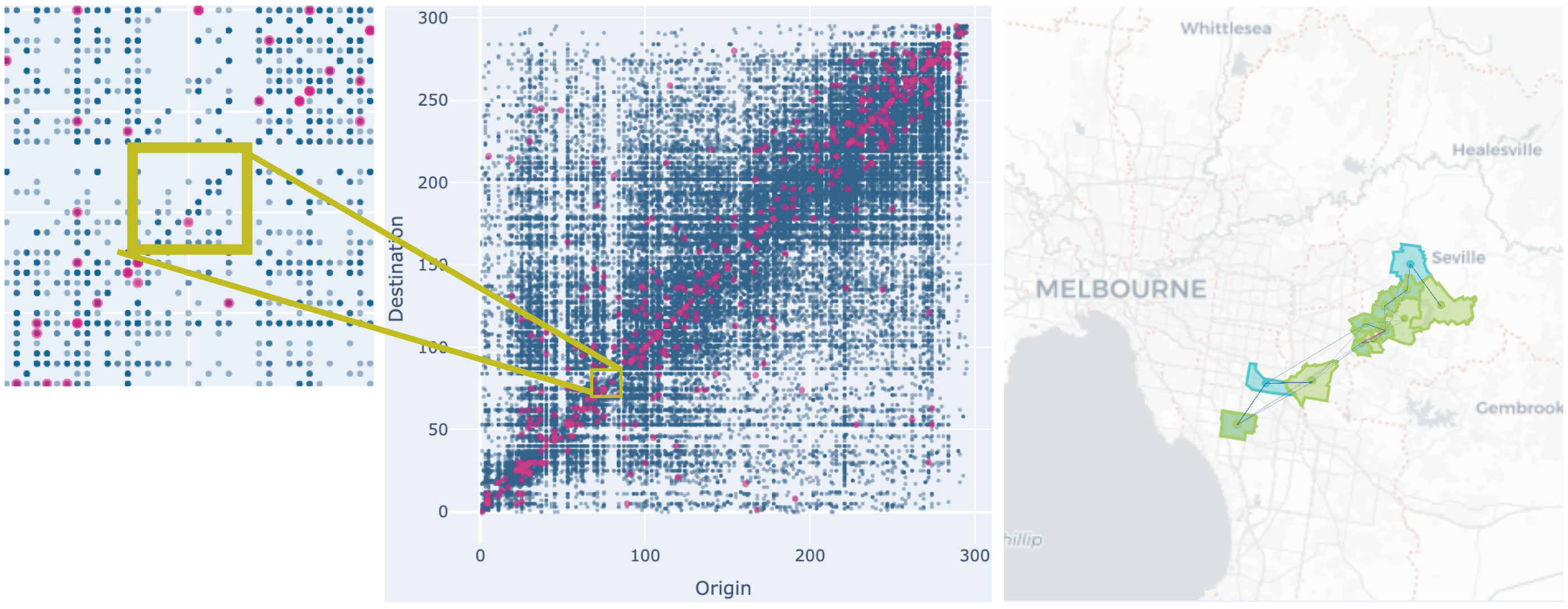}
    \caption{\textcolor{blue}{Case Study \ref{subsec:caseA} (A): Short-distance trip patterns. A cluster (green rectangle) is selected in the OD-Plot View (left) with corresponding locations shown in the Trip Location View (right).}}
    \vspace{-0.3cm}
    \label{fig:casestudy1_example1}
\end{figure}

Several local patterns can also be identified in the OD-Plot. For instance, the bottom-left cluster highlighted in \figref{casestudy1_example1} indicates a city region where most accidents occur in short-distance trips. By selecting this cluster with the rectangular tool in the OD-Plot, the corresponding region can be visualized in the Trip Location View. The analysis reveals that these trips are concentrated in a peripheral eastern area of the city and can be characterized as local traffic. Only a single fatal accident (purple) is observed in this cluster.  

\begin{figure}[!t]
    \centering
    \includegraphics[width=1.0\columnwidth]{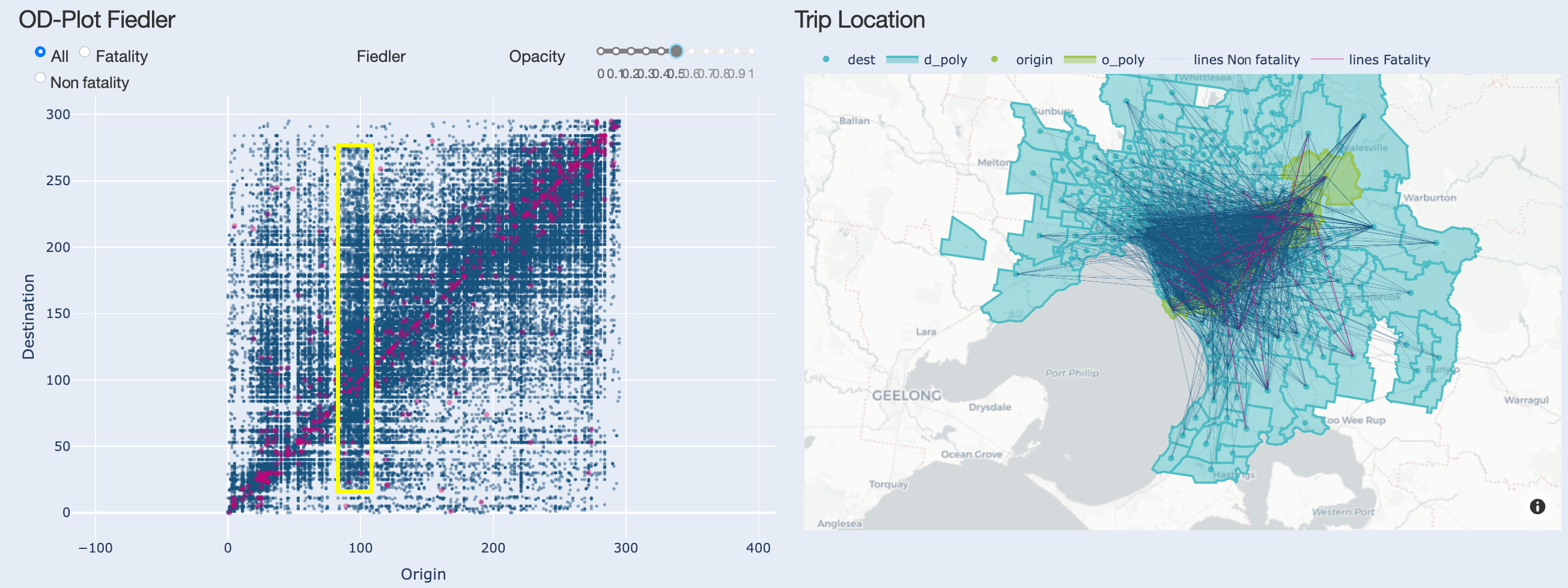}
    \caption{\textcolor{blue}{Case Study \ref{subsec:caseA} (A): Vertical subset of trips with varying lengths selected in the OD-Plot View (left). The Trip Location View (right) shows that these trips originate mainly in downtown Melbourne and spread across different areas. Fatal accidents (purple) are relatively concentrated in short trips, while non-fatal accidents (blue) are more broadly distributed.}}
    \label{fig:casestudy1_example2}
    \vspace{-0.7cm}
\end{figure}

\figref{casestudy1_example2} illustrates an opposite situation, where a subset of trips is selected on the OD-Plot. The selected points are arranged vertically and include trips with various lengths. The Trip Location View reveals that the origins are concentrated in the downtown area of Melbourne, with trips flowing to diverse areas. Fatal accidents (depicted in purple) tend to occur during shorter trips, possibly due to the heavy traffic in downtown Melbourne. 

\begin{figure}[!t]
    \centering
    \includegraphics[width=1.0\columnwidth]{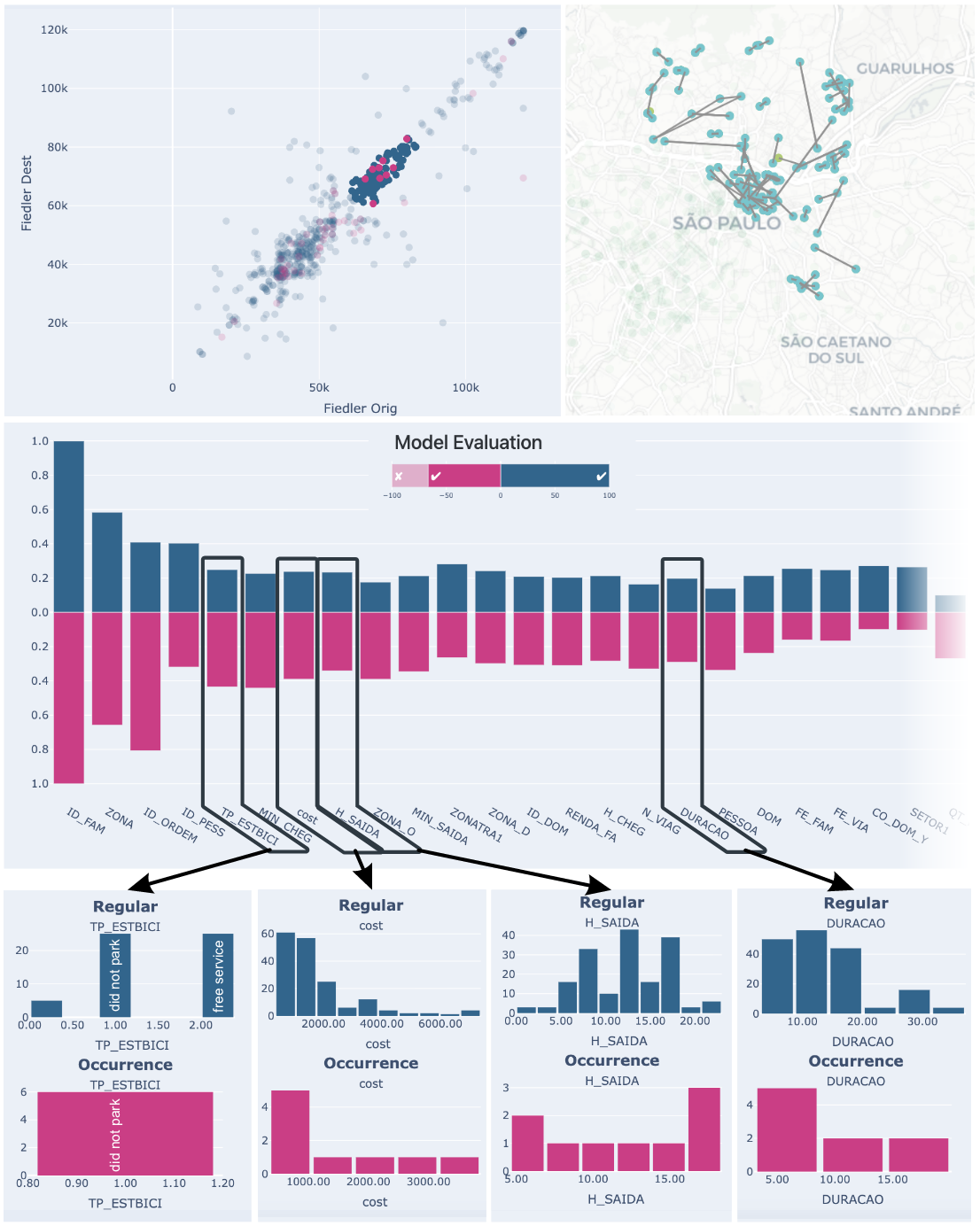}
    \caption{\textcolor{blue}{Case Study \ref{subsec:caseA} (B): OD-Plot of bicycle trips in São Paulo. The figure highlights the predominance of short trips (points along the diagonal) and the scarcity of long-distance trips, which is consistent with bicycle transportation. Case Study \ref{subsec:caseB} (B): The second row shows important features, expanded in the third column. \textcolor{blue}{The blue plots exhibit patterns that are largely opposite to the purple plots, representing trips with theft, particularly for attributes such as parking type, cost, departure time, and trip duration. Case Study \ref{subsec:caseC} (B): Accuracy favors non-robbery trips.}
}}
    \label{fig:casestudySaoPaulo}
    \vspace{-0.7cm}
\end{figure}

\textcolor{blue}{\textbf{Dataset B.} \figref{casestudySaoPaulo} presents the OD-Plot of bicycle theft occurrences in São Paulo. At first glance, the dataset appears highly imbalanced, with a majority of trips without theft (blue). The concentration of points near the diagonal highlights the predominance of short-distance trips, consistent with the nature of bicycle mobility. In contrast to Dataset A, long-distance trips (far from the diagonal) are scarce.} \textcolor{blue}{The selection mechanism further illustrates how the chosen region becomes more opaque, while non-selected points decrease in opacity, making overlaps more perceptible. In the Trip Location View, origins and destinations are represented as individual points, reflecting the finer granularity of Dataset B, in contrast to the polygon-based regions of Dataset A.}  

\textcolor{blue}{Overall, these case studies demonstrate that \ordena enables experts to interactively filter trips with simple selection tools and examine their spatial distribution (task T1). Moreover, both global and local OD patterns can be effectively identified and interpreted through the linked OD-Plot and Trip Location Views (task T2).}  

\subsection{Understanding Patterns in \textcolor{blue}{Trips}}
\label{subsec:caseB}
This case study aims to visually identify and compare spatial patterns in \textcolor{blue}{trips} with and without \textcolor{blue}{associated incidents}, relating these patterns to \textcolor{blue}{their corresponding} attributes. For the latter, we leverage the explainability resources provided by \ordena.  

\textcolor{blue}{\textbf{Dataset A.}} \subfigref{casestudy2-odplot2.pdf} shows the OD-Plot of accidents with fatalities (with the “Fatality” option activated in the interface). One can observe that, in most areas, such accidents are concentrated near the diagonal, indicating that fatalities frequently occur in short-distance trips.  

Other noteworthy patterns revealed in \textcolor{blue}{\subfigref{casestudy2-odplot2.pdf}} are the voids along the diagonal around indices 80, 160, and 200 (highlighted by squares). These voids indicate that trips starting and ending in these regions rarely result in fatalities, despite the fact that these areas register a high number of accidents overall, as seen in \subfigref{casestudy2-odplot1}. Particularly relevant are the cases near index 80, where trips are very short. By contrast, accidents near index 200 are more broadly distributed and predominantly non-fatal, whereas index 160 exhibits an intermediate behavior.  

As illustrated, \ordena makes it possible to uncover regions where short-distance trips exhibit markedly different fatality rates, a phenomenon that would be difficult to detect without the OD-Plot. In this sense, the OD-Plot directly supports task T2 and partially contributes to task T4.  

\subimages[!t]{\textcolor{blue}{Case Study \ref{subsec:caseB} (A):} OD-Plot visualization of accidents. The yellow squares indicate regions with short-distance trips that \textcolor{blue}{involve several accidents (b)} but rarely result in fatalities \textcolor{blue}{(a)}.}{casestudy2-odplot}{
\subimage[Accidents with fatality.]{0.495}{casestudy2-odplot2.pdf}
\subimage[Accidents without fatality.]{0.5}{casestudy2-odplot1}}

\begin{figure}[!t]
    \centering
    \includegraphics[width=1.0\columnwidth]{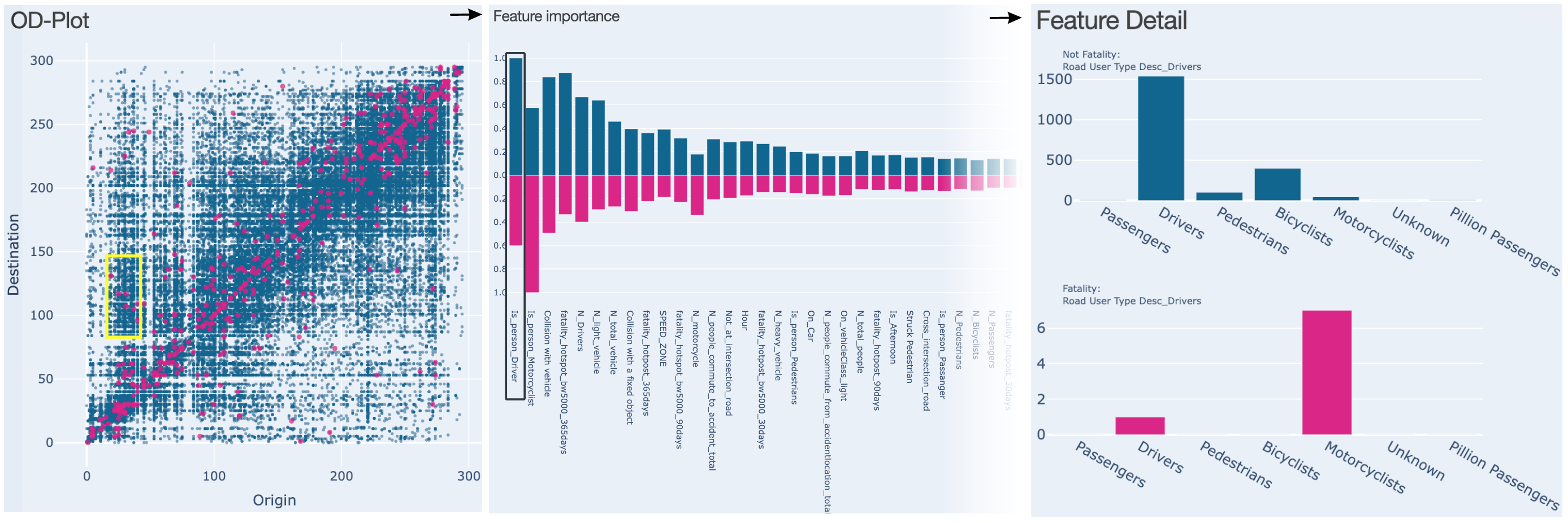}
    \caption{\textcolor{blue}{Case Study \ref{subsec:caseB} (A): Analysis of a cluster of short-distance trips (yellow rectangle in the OD-Plot). The Feature Importance View (middle) highlights the most relevant attributes. The Feature Detail View (right) shows that motorcyclists are the most common fatal victims (bottom), while car drivers are the group most likely to survive accidents (top).}}
    \vspace{-0.7cm}
    \label{fig:casestudy2_example1}
\end{figure}

\ordena also provides mechanisms to analyze and compare the feature importance of selected subsets of accidents. When users select a subset of points in the OD-Plot, the Feature Importance View is dynamically updated, as illustrated in \figref{casestudy2_example1}, where a cluster of trips with and without fatalities is highlighted. The Feature Importance View identifies ``Is\_person\_Driver'' as the most relevant feature. By clicking on this feature, the Feature Detail View (right panel in \figref{casestudy2_example1}) displays the statistical distribution of its values for fatal and non-fatal accidents (task T3). The purple distribution shows trips with fatalities, revealing that motorcyclists are the most common fatal victims. Conversely, the blue distribution indicates that survivors are most frequently car drivers.  

\begin{figure}[!h]
    \centering
    \includegraphics[width=1.0\columnwidth]{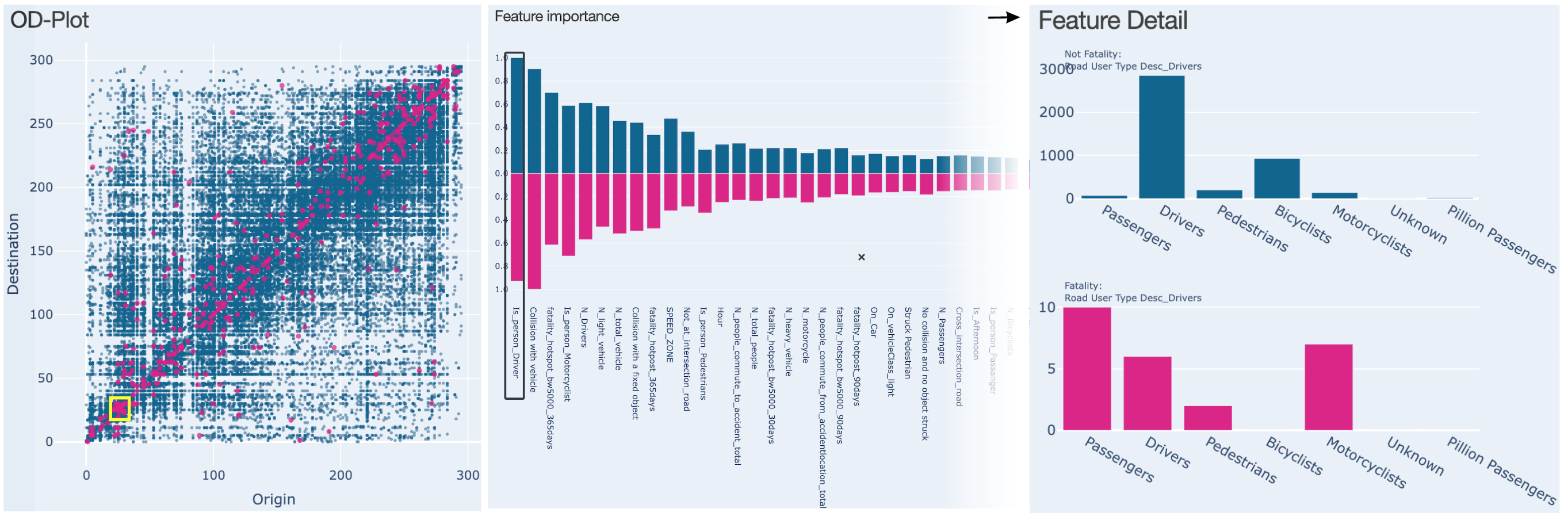}
    \caption{\textcolor{blue}{Case Study \ref{subsec:caseB} (A):} Analysis of another cluster of short-distance trips. The Feature Importance View (middle) again identifies ``Is\_person\_Driver'' as the most relevant feature, but the distribution differs substantially from \figref{casestudy2_example1}, revealing passengers as the primary fatal victims. This finding is corroborated by the second most relevant feature, which indicates that many fatal victims are pedestrians.}
    \label{fig:casestudy2_example2}
\end{figure}

Analyzing another cluster of short-distance trips (\figref{casestudy2_example2}), ``Is\_person\_Driver'' remains the most critical feature. However, the distribution of this attribute (right panel) reveals a distinct pattern compared to the previous case: passengers constitute the majority of fatal victims, surpassing not only motorcyclists but also drivers, who remain the most frequent group among non-fatal accidents.  

\textcolor{blue}{\textbf{Dataset B.} \figref{casestudySaoPaulo} illustrates a selection that includes both bicycle trips with and without theft. For demonstration purposes, we focus on a subset of intuitive features that also rank among the top positions in the Feature Importance View. The bottom row of the figure expands these selections. For example, regarding parking type, most trips without theft involved ``did not park'' or ``free service,'' whereas most stolen bicycles were associated with trips in which no parking occurred, suggesting negligence by owners or theft during transit. A second attribute is cost: although most bicycles in the dataset cost less than 2000 BRL, a few high-end models appear. Interestingly, stolen bicycles are predominantly from the lower price range, which may suggest that expensive bicycles are more carefully protected by their owners and less exposed to risky routes. A third attribute is departure time: regular trips are concentrated in peak hours (8 a.m., noon, and 7 p.m.), while thefts occur mostly at night or in the early morning. Finally, the attribute of trip duration indicates that most trips last less than 20 minutes, but thefts are heavily concentrated in shorter trips (5–10 minutes), likely reflecting journeys interrupted by the criminal event. We selected these features (parking type, cost, departure time, and duration) not only because they are ranked highly in the Feature Importance View, but also because they are readily interpretable by domain experts. This choice increases the transparency and credibility of the analysis, ensuring that the observed patterns are both statistically relevant and practically meaningful.}  

\textcolor{blue}{These cases} demonstrate that the analytical resources of \ordena effectively reveal patterns in accidents with and without \textcolor{blue}{incidents} (T4), while also facilitating the visual examination of relevant attributes and their statistical distributions (T3).  

\subsection{Exploring the predictive model on different regions}
\label{subsec:caseC}
This last case study focuses on assessing the performance of a predictive model, considering only a subset of accidents selected from the OD Plot View. The goal is to evaluate whether the models perform differently depending on the trips' origin and/or destination, figure out how the features may influence this performance, and determine the reliability of the feature importance evaluation.

\textcolor{blue}{\textbf{Dataset A.}}
Consider the subset of accidents chosen in \figref{casestudy2_example1}.  The Model Evaluation View, depicted in \subfigref{casestudy3-modeleval1}, shows the prediction performance of the model on the selected subset. It displays a good predictive performance for both classes, with a small rate of false positives, that is, accidents without fatalities classified as those with fatalities (light blue rectangle on the right with an x inside) and no false negatives.
Differently, \subfigref{casestudy3-modeleval2} reveals that, in the subset from \figref{casestudy2_example2}, the model resulted in a larger number of false positives when compared against the subset of trips selected in \figref{casestudy2_example1}, and a small number of false negatives (light red rectangle on the left with an x inside). 

\subimages[!t]{\textcolor{blue}{Case Study \ref{subsec:caseC} (A):} Model performance on two different subsets of accidents.}{casestudy3_modelevals}{
\subimage[Accidents selected in \figref{casestudy2_example1}.]{0.485}{casestudy3-modeleval1}
\subimage[Accidents selected in \figref{casestudy2_example2}.]{0.485}{casestudy3-modeleval2}}

We can examine the Feature Detail View of \figref{casestudy2_example1} to gain deeper insights into model behavior. This view shows the distribution of the most important features, revealing a clear pattern: motorcyclists are the most common fatal victims. However, \figref{casestudy2_example2} presents a contrasting picture. Here, the distribution of fatal victims lacks a clear trend, with a significant number of passengers, drivers, and motorcyclists involved. This lack of a dominant pattern in the second case might contribute to model misclassifications of some accidents as fatalities.

\textcolor{blue}{\textbf{Dataset B.} \figref{casestudySaoPaulo} shows the model evaluation view, where higher accuracy is achieved for the class of non-robbery trips. This may be explained by the underrepresentation of robbery trips in the analyzed sample.}

This case study attests that the visual analytic mechanisms enabled by \ordena can support tasks T4 and T5, providing resources to assess the model performance and reasoning about the reasons behind the observed behavior.

%% file: 7.expert_evaluation.tex
\section{Expert Evaluation}

This section describes the expert evaluation we have accomplished, focusing mainly on getting detailed feedback about \ordena\!\!’s Utility, Usability, and Novelty.

\ordena was presented to four experts in Origin-Destination data analysts not involved in its development through an illustrative video and a live demonstration of its functionalities. Subsequently, we conducted guided interviews. \textcolor{blue}{Participants were also granted direct access to the deployed web tool \url{https://dash.giva.icmc.usp.br/}, allowing them to independently explore and interact with the system at their convenience} The experts have over 15 years of experience in this domain and hold PhDs. For a detailed description of the experts, the questions asked, and the provided answers, please refer to the \textit{Supplementary Material}.

\myparagraph{General impressions:} The experts rated \ordena an average of 4 out of 5. They primarily appreciated the tool's dynamic capabilities for exploring OD data and integrating techniques to explain the model within the tool. 

\myparagraph{Utility:} All the experts agreed that \ordena is useful, particularly highlighting the tool's OD-plot selection, which converts to trip location, and the Feature Importance View as the most valuable components. However, one expert suggested that including informational pop-ups would make the tool more useful to non-specialized practitioners, such as policymakers.

\myparagraph{Novelty:} While most experts had previously conducted analyses similar to those presented in the case studies, they noted that the functionalities enabled by \ordena are unique among OD visualization tools. Experts familiar with similar analyses using other tools mentioned R and Python, whereas those who had only partially performed similar analyses cited ArcGIS\footnote{https://www.arcgis.com/index.html} and TransCAD\footnote{https://www.caliper.com/tcovu.htm}. This corroborates with the level of novelty incorporated into \ordena for exploring OD data, particularly for non-programmers.

\myparagraph{Usability:} Regarding the usability of \ordena\!\!, opinions among the experts were generally positive but varied slightly. Half of the experts described the interface and design of the tool as intuitive and easy to use, while others found it to be partially intuitive. When compared to other tools, three out of four experts agreed that \ordena offers better usability than those they have previously used. The fourth expert had no comparable tool experience for a direct comparison. Additionally, one expert highlighted \ordena\!\!’s potential to deliver faster results with less effort.

\myparagraph{Suggestions:} The experts recommended several enhancements to \ordena that they considered important. These include displaying distributions across time and space, changing the predictive model within \ordena\!\!, and speeding up the update process for the Feature Importance View. They also suggested including urban planning data, allowing the selection of regions of interest in the Trip Location View, and altering the line representation of trips in the Trip Location View to edge bundling. Furthermore, they advised replacing the numerical labels on the OD plot axes with neighborhood names for clearer referencing, adding pop-up information to explain the elements and methodological aspects of \ordena and integrating feature importance data directly into the map.

%% file: 8.discussion.tex
\section{Discussion and limitations}
The case studies discussed in \secref{cases} show that \ordena can successfully tackle the \textcolor{blue}{requirements and} tasks described in \secref{requirements} through a simple, intuitive, and easy-to-use visualization tool.
In particular, the OD-plot is capable of unveiling patterns that are difficult to identify with other commonly used visual metaphors such as OD-matrix and flow maps.


An important aspect not taken into account in our implementation is that the Fiedler vector may present an anisotropic behavior, making not-so-close locations be placed close to each other on the OD-plot axes. The anisotropy can be controlled/mitigated by properly tuning the edge weights on the Laplacian matrix~\cite{berger2010fiedler,ding2001min}. How to tune edge weights to further improve OD-plot is an interesting subject for follow-up work.     
\textcolor{blue}{The OD-plot should be interpreted as a structural view of mobility rather than a geographically faithful representation. While this abstraction reveals latent hierarchies and connectivity-based patterns, the findings should not be interpreted as causal but rather as descriptive signals that support hypothesis formulation. Previous work \cite{salinas2024visual} has explored alternative orderings such as t-SNE, UMAP, and other spectral approaches, and our method complements these efforts by showing how Fiedler-based ordering uncovers multi-scale connectivity structures.}

We intend to apply \ordena to analyze trips in cities with different characteristics. \textcolor{blue}{Several} interesting questions can be investigated in this context. For instance, are patterns such as regions with a large number of short-distance trips \textcolor{blue}{involving incidents} also present in other cities? Are the most important attributes \textcolor{blue}{consistent across} different cities? \textcolor{blue}{These} and other questions will be \textcolor{blue}{addressed in future work}.

It is important to emphasize that the context of traffic accidents \textcolor{blue}{and bike theft represent only a few application scenarios} for the proposed framework. \ordena possesses versatility and applicability across a spectrum of tasks. In fact, the proposed framework is well-suited for any application involving movement data with geolocated origin-destination points, associated attributes\textcolor{blue}{,} and a target classification task. Examples of such applications include ride-hailing app transportation with crime events, investigation of shipping activities involving loss of merchandise, and exploration of package delivery services where mistakes are probable. These examples highlight the wide-ranging potential applications \textcolor{blue}{of} \ordena.

%% file: 9.conclusion.tex
\section{Conclusion}
In this work, we proposed \ordena\!\!, a visual analytic tool to support the analysis of transport data. We developed \ordena in close collaboration with domain experts, translating their analytical needs into the visualization system. We also introduced the OD-plot, a scatter plot-based mechanism to represent OD-flows. The OD-plot naturally reveals different spatial patterns while enabling an intuitive interactive mechanism to select and filter subsets of OD flows. \ordena was validated through three case studies using real data, showing its effectiveness in tackling the tasks pointed out by the domain experts.